# The group height of spicules links their acceleration and velocity

Leonard A Freeman  23 Hope Street, Cambridge, CB1 3NA, UK.
 leonard.freeman@ntlworld.com           Orcid  0000-0002-1242-0892

**Abstract**

This study reveals a new feature of many solar jets: a group height, which links their acceleration and velocity.

The acceleration and velocity *(a,V)* for jets such as spicules, often displayed as scattergraphs, show a strong correlation. This can be represented empirically by the equation, $V = pa + q$, where *p* and *q* are two arbitrary non-zero constants.

This study reanalyses the *(a,V)* data for nine different groups of jets, in order to test an alternative proposal that a simpler relationship directly links *(a,V)* to the mean height for the group of jets, without needing the empirical constants *p* and *q*. A standard mathematical test – plotting *log(a)* against *log(V)*, tests whether $V \sim a^n$ and if so, gives the value of n. When this is done for a wide range of jets the index *n* is consistently found to be close to 0.5

The nine groups of jets include spicules, macrospicules and dynamic fibrils. The result, $V \sim a^{0.5}$, or equivalently $V^2 = ka$, with only one constant, provides as close a match to the data as the equation $V = pa + q$, which requires two unknown constants. It is found that the constant *k*, is a known quantity: just twice the mean height, $\bar{s}$, of the group of jets being analysed. This then gives the equation $V^2 = 2\,a\,\bar{s}$, for the jets in the group. This more succinct relationship links the acceleration and maximum velocity of every jet in the group to a well-defined quantity - the mean height of the group of spicules, without needing extra constants.

## 1. Introduction

There are millions of needle-like jets of plasma, known as spicules, that continually rise and fall all over the sun's surface. Their behaviour has been recorded by a number of researchers ,eg. Duan Y. et al., (2023);  Lippincott, (1957); De Pontieu et al. (2007a,b); Pereira, De Pontieu & Carlsson, (2012); Zhang et al (2017); Rouppe van der Voort et al. (2007); Loboda & Bogachev, (2017, 2019).  A comprehensive review is given by Tsiropoula et al (2012).

The motion of an individual jet begins when it is ejected from the solar surface with a maximum velocity, V. It then rises up, undergoing a constant deceleration, a, until it reaches its maximum height, s. It then returns, accelerating back down to the surface, where it regains its maximum speed, V. Although this is very similar to ballistic behaviour due to gravity, it is considered that this is not a gravitational effect: spicule acceleration is unlike that due to gravity. For example Heggland et al (2007) found a range of jet decelerations, and even allowing for the inclination of jets to the vertical, concluded that " *this distribution is extremely hard to reconcile with a ballistic model, and indicates that we should look not to gravity but elsewhere in trying to explain it*."   Rouppe van der Voort et al, (2007), in their observation of quiet sun mottles, state that their "*deceleration is not related to gravity, but completely determined by shock wave physics*". A similar view is expressed by De Pontieu et al (2007a). Figure 1 illustrates this difference, showing how the acceleration – velocity relationship for the jets observed by Duan et al (2023) compares with ballistic behaviour at the surface of the sun, where the deceleration , g is 274 ms$^{-2}$ and of course, constant. Ballistic motion can be imagined if we consider a group of objects thrown upwards with random velocities: this launch velocity is completely independent of the subsequent deceleration. However for the solar jets there clearly is a relationship between acceleration and velocity. The aim of this paper is to investigate this "*most striking correlation*" (De Pontieu 2007a) and see what it can tell us.

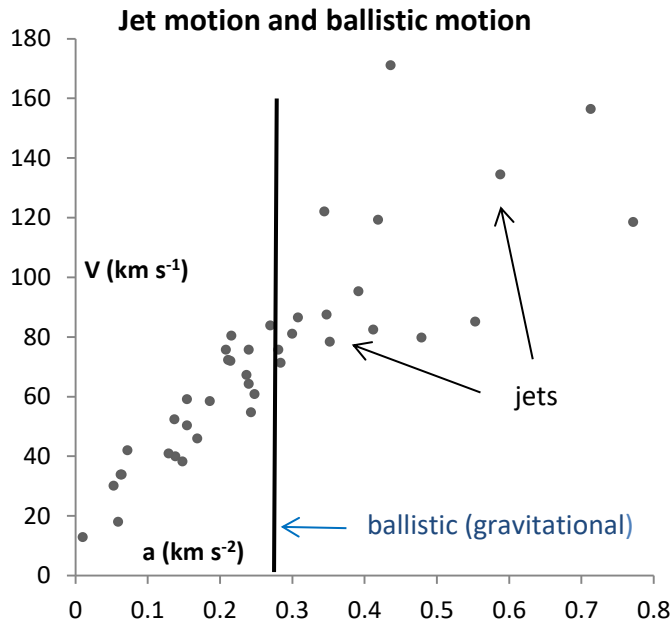

**Figure 1**. The jets (black dots) show a wide range of accelerations, unlike ballistic motion where the acceleration is constant (shown by vertical line). For ballistic motion the (a,V) correlation is zero, but the jets show a high correlation, up to ~ 0.9.

This scattergraph, and the others have been recreated by digitisation of the data

Loboda et al (2017) measured the kinematic properties of a group of 15 macrospicules. Freeman (2019) suggested a new relationship between their maximum velocity $V$, and acceleration, $a$, namely $V^2 \sim a$.

Here, a more wide ranging set of data from different types of jet is undertaken to test the validity and scope of this proposed relationship. And as a result of this analysis, the importance of the group height of spicules becomes apparent.

Measurement of the height against time for a single jet normally results in a parabola, indicating a constant deceleration. For a single spicule, with its constant deceleration the two equations connecting the variables (a,V) are:

$$V = a\,t \qquad (1)$$
$$V^2 = 2\,a\,s \qquad (2)$$

where $t$ is the time taken for the spicule to reach its maximum height, $s$. These four kinematic variables can all be found from the height–time parabola.

From *(a,V)* scattergraphs such as shown in figure 1, a trend line through the points is generally found, $V = pa + q$, where p and q are two empirical constants. The line found by Duan et al (2023), for the data of figure 1 was:

$$V = 170a + 26 \qquad (3)$$

Figure 2 shows this trend line and the data points.

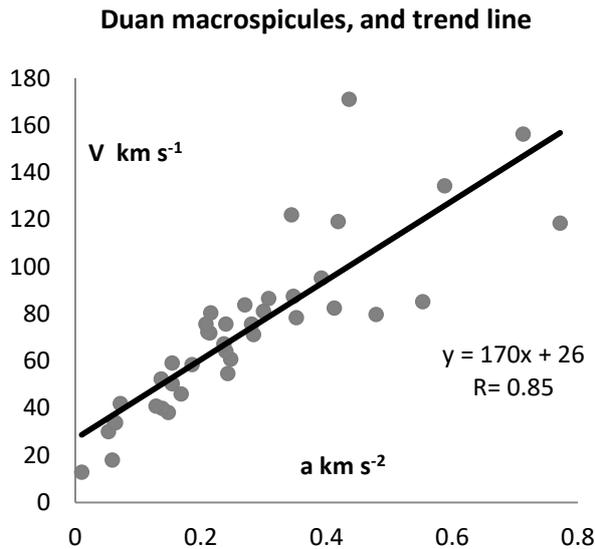

**Figure 2.** Showing the data points for the macrospicules of Duan et al (2023), and their trend line. The correlation of the points to the trend line is 0.85.

## 2. Testing the mathematical relationship

The alternative relationship between *(a,V)*, proposed by Freeman (2019) is a square law, $V^2 \sim a$, similar to equation (2). So for an *(a,V)* scattergraph this becomes $V = ka^{1/2}$. If we need to test a relationship such as $V = ka^n$, the standard method is to plot a graph of *log(V)* against *log(a)* for the data points: the gradient gives the value of *n*. This provides a simple way of checking: for proportionality between *(a,V)*, *n = 1*, whereas for the square law $n = 0.5$ ( $V^2 \sim a$, so $V \sim a^{0.5}$ ). This process has been carried out for a variety of jets, to test the proposal.

The result for the data in figure 1 reveals *n = 0.60*. This trend curve is shown in figure 3.

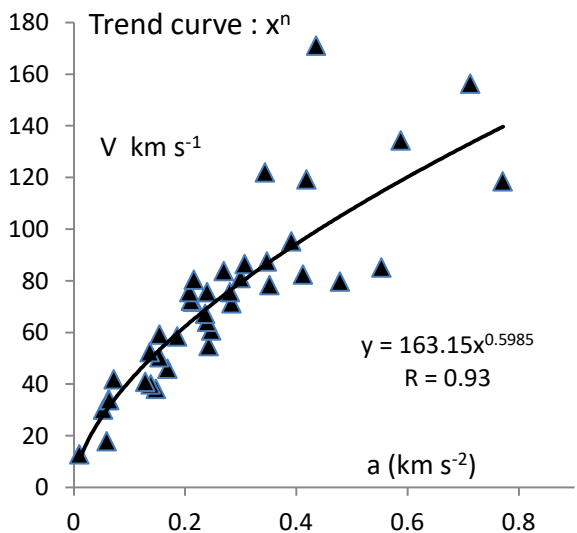

**Figure 3.** Showing the optimum curve of form $x^n$ for the data points of Duan et al (2023)

The main conclusions are:

1. The result is closer to the square law $V = ka^{0.5}$ than a proportional law, $V = ka$.

.2. The correlation coefficient is slightly better for the curve (0.93) than for the trend line (0.85) of figure 2.

Figure 4 shows some results for dynamic fibrils by De Pontieu et al (2007a), from their region 1 (blue diamonds). In this case the trend line is V = 68a + 9.5

The optimum curve can also be seen in the figure. Its equation, $V = 46a^{0.46}$, again shows a relationship close to $V \sim a^{0.5}$.

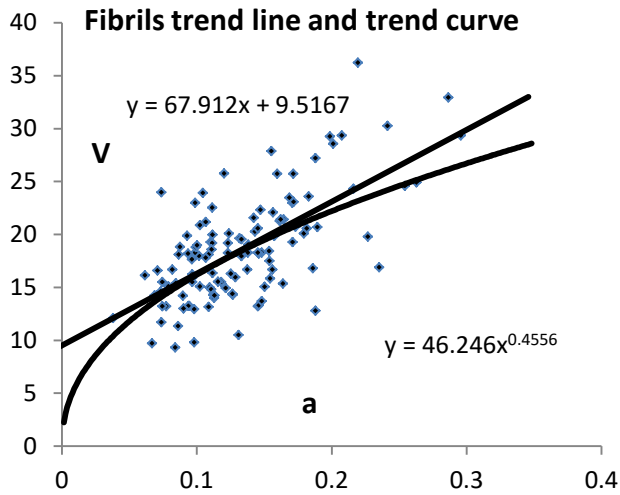

**Figure 4.** The data points for the dynamic fibrils (De Pontieu et al, 2007a) with the trend line and optimum curve.

### 3. Survey of data on groups of jets

The (a,V) relationship has been tested for a wide range of jets, such as spicules, mottles, fibrils and light bridges, and the results for the value of "n" for each group is given in table (1).

**Table 1        Value of index 'n' for different jets**

| Type of jet | n if $V \approx a^n$ |
|---|---|
| Macrospicules, Duan 2023 | 0.60 |
| Dynamic fibrils, region1, De Pontieu et al 2007a | 0.46 |
| Quiet Sun mottles 37points, De Pontieu et al 2007b | 0.43 |
| Sunspot oscillations, Tian 2014 ApJ | 0.52 |
| Light bridges , Zhang J., et al 2017 | 0.40 |
| Macrospicules, Loboda et al 2017 | 0.54 |
| QS Mottles Rouppe et al  2007 | 0.46 |
| Quiet Sun mottles (blue triangles), De Pontieu et al PASJ 2007 | 0.49 |
| Type 1 spicules (red triangles), De Pontieu et al, PASJ 2007 | 0.51 |
| **Mean value of n** | **0.49±0.06** |

It's clear from Table 1 that data from all the different jets and samples strongly indicate that $V \sim a^{0.5}$, or $V^2 \sim a$. This rules out a proportional relationship, (for which n=1), as suggested for instance by Zhang et al (2012) and Anan et al (2010).

It also means that, over the data range considered, there are two equations which are equally good in representing the empirical data for each spicule group:

$V = pa + q$  or  $V^2 = ka$, where p, q and k are constants.

## 4. The trend line and the curve

### 4.1 Problems with the line
Loboda et al, (2019), examined a number of studies and found wide variation of the constants *p* and *q* in the linear equation. They found that the non-zero intercept q presented a particular difficulty in trying to apply the shock wave theory to explain spicule behaviour.

Of course any curve can be represented by a line over a short section of the curve, but caution is needed is needed if extrapolating outside the data region: the straight line equation may not then be physically meaningful. For example, consider points at or very close to the intercept, (a,V) = (0,26), as in Figure 2. This would represent jets with a constant upward velocity of 26 kms$^{-1}$, and with zero deceleration. Such jets have not been recorded in the studies referred to here. And if the trend line is continued to negative values of acceleration, it is hard to imagine what this could physically represent. The alternative choice of $V^2 \sim a$ does not have these problems, since its origin is always (0,0): the troublesome intercept q disappears.

### 4.2 Identifying the constant *k* in $V^2 = ka$
We can test the suggestion that the constant *k* is simply twice the mean height of the spicules, by writing the equation as:

$$V^2 = 2\,a\,\bar{s} \qquad (4)$$

or $\quad V = (2\,a\,\bar{s})^{0.5} \qquad (5)$

Where $\bar{s}$ is the mean height of the group of spicules.

The height of an individual jet can either be measured directly, or found from its *(a,V)* values, using $s = V^2/(2a)$. The mean height can then be calculated. Referring back to the spicule data of Duan et al, in figure (1), this height, $\bar{s}$ is found to be 10700 km, so $2\bar{s} = 21400$ km. Figure 5 shows this value, substituted into equation (5). For comparison the trend line is also shown.

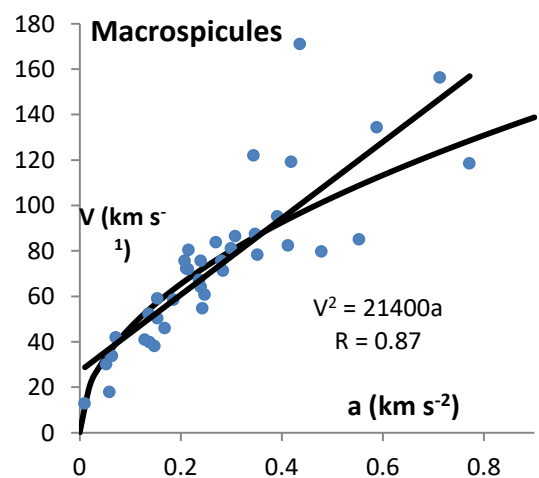

**Figure 5** The graph of $V^2 = 21400a$ and the trend line for comparison. The correlation coefficient for the curve is 0.87, slightly higher than the trend line.

A similar picture emerges for other jets, such as the dynamic fibrils of figure (4). This is shown in figure (6). In this case the mean height of the fibril group is 1400 km, so 2s=2800.

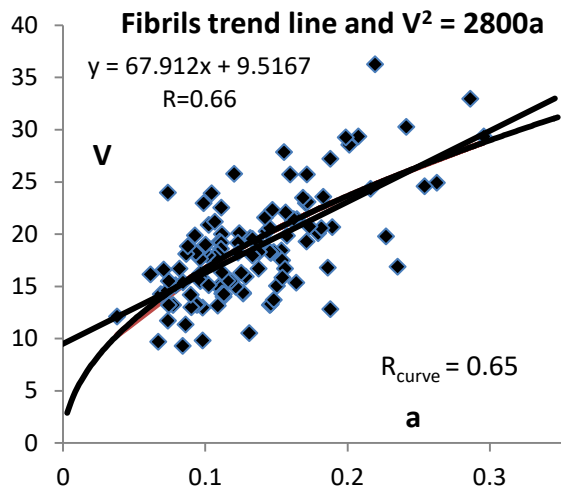

**Figure 6.** This shows that the alternative equation (4), $V^2 = 2\,a\,\bar{s}$ gives approximately the same correlation for the dynamic fibrils as the linear equation.

The alternative equation:
1. Removes the problem of the non-zero intercept.
2. Does not require two empirical constants.
3. Is physically meaningful. If the height of all spicules were exactly the same, all the data points would be on a single curve, The closeness of the correlation to 1.0 tells us how closely the points approach the curve.

## 5. Importance of groups

Note that the high correlations given by both the curve and the trend line only come from a single group of spicules. If we merged all the data for all types of spicules in different regions, then we would have a much greater spread of points, and the (a,V) correlation would become so low as to be of little use. This is of course recognised implicitly in the studies referenced here. Pereira et al, 2012, for example investigated the differences in three regions: coronal holes, quiet sun and active regions. Amongst the differences, coronal holes had the longest spicules. This was attributed to different magnetic field configurations.

This last point means that in order to get strong (a,V) correlations we may need to work with an area with a stable magnetic field.

The group behaviour of spicules has long been recognised. For example, Lippincott, (1957) described a common pattern of the jets as a *"wheat field"* of *"parallel spicules with approximately the same heights"* over an average length of 140,000 km. She also found typical lifetimes of around five minutes and noted that spicules were *"growing and disappearing as a unit"*.

Skogsrud et al (2014) report: *"it is apparent that many spicules do not evolve independently, but rather evolve as small groups of spicules in a collective fashion. The spicules in the groups appear and disappear nearly at the same time and have similar apparent motion."*

It can be seen that the new group equation identifies the characteristic height of a spicule group with the constant $\bar{s}$ in the equation . Because of the high (*a,V*) correlation this constant is more than just a mean value: any set of numbers has a mean, but the correlation allows us to think of $\bar{s}$ as a characteristic, or typical height of the group of spicules. The closer the correlation comes to the perfect value of 1.0, the closer the scattergraph is to the single curve given by $V^2 = 2\,a\,\bar{s}$, which represents jets of exactly the same height.

## 6. Observational methods

Measuring the properties of solar jets is not easy. They lie at the limit of resolution, are faint and often overlapping, in the "forest" of spicules. The individual accuracy for some of the more difficult jets, quiet sun mottles, Rouppe van der Voort et al (2007) estimate at 10%.

Solar jets are generally observed at the wavelength of Hα (657nm) or Ca II H (397 nm) lines. It's important to investigate whether the choice of the line affects the measurements of spicule lengths or other properties. Simultaneous measurements with these two wavelengths were made by McIntosh et al (2008). They found that these methods showed similar heights and dynamic behaviour of spicules. Rouppe van der Voort et al (2013) examined short dynamic fibrils, and also found "*a close correspondence .....with a small offset of Hα being slightly taller than Ca II* ". Pereira et al (2012) compared spicule properties measured in different lines and from different areas. They found differences for spicules velocities and lifetimes, but reasonable agreement for lengths and inclinations. In the studies examined here, other shorter wavelengths have also been used, eg Si IV and He II, the shortest wavelength, (30.4nm). Using different lines is helpful in determining other jet properties such as temperature and density.

## 7. Discussion and Summary

Given that there are probably slight differences in measurements made using different emission lines, for the groups in Table 1, it is encouraging to note that the results are reasonably consistent with each other in producing the relationship $V \sim a^{0.5}$. The jets in these groups also show a common parabolic space-time pattern. Since measurements are made with just one method normally, it seems likely that we are seeing the real general behaviour of jets, although there may be slightly different values for the mean heights, for example, from different observational filters. And there may be greater causes for observed differences: Pereira et al (2012) suggest that many differences are mediated by regional differences in the magnetic field configuration.

The commonly used trend line equation gives a good representation of the connection between acceleration and velocity, but requires two empirical constants, whose origin was unexplained. The results here show how a new relationship simply links *(a,V)* with the mean height of the group of spicules, $\bar{s}$, without the need for two undefined constants.

The *(a,V)* correlation indicates how close the scattergraph points come to the single curve given by equation (5). It has generally been known that jet heights vary with the region from which they emanate, and this analysis provides a mathematical confirmation for the existence of a characteristic height for a group.

The mathematical relationship between the kinematic variables does describe how these jets move, but doesn't give us the reason why, or the underlying physics. However, if we know the relationship it provides a test for the theory: does it predict the motion of jets? The linear equation $V = pa + q$ has been used to test the theory that jets are shock-wave driven: eg Heggland et al (2007). But Loboda et al (2019) found difficulties in reconciling the non-zero constant *q* with the theory. Perhaps the alternative equation derived here may be of use here, if the shock wave theory could be modified.

The underlying physics of these jets remains to be decided. Two contenders for their origin are the theory of magnetic reconnection, where rapid changes in the magnetic field occur or shock waves. The analysis here brings another focus – onto the characteristic height, or length of a jet group: what is controlling the height? The general consensus is that spicules are strongly influenced by magnetic fields – for instance Kuridze et al (2024) conclude that: "morphological characteristics, such as orientation, inclination, and the length of fibrils, are defined by the topology of the magnetic field" This raises the possibility that the physical scale of the magnetic field itself may be involved.

The scale height, or more accurately the scale length of local fields (given that both fields and jets are usually inclined to the vertical) on the sun can be defined mathematically by a decreasing exponential or an inverse square law, such as $B(z)/B_0 = (z_0/z)^2$, where $z_o$ represents the scale length. At an origin $z = z_0$, $B(z)/B_0 = 1$, but at $z = 2z_0$, $B(z)/B_0 = ¼$. So the scale length gives an indication of the physical extent of the magnetic field. Morton et al (2012) found that the magnetic field along a solar jet decreased by half after 2000 km approximately. Schad et al (2013), examined the fields of some fibril groups. For three out of the four fibril fields they show that the magnetic field typically falls 70% or more by the end of the fibril. This, and other studies, indicate that the magnetic scale length is generally of the same order as the length of the jet itself. If this is correct, the height of jets could be a proxy for the magnetic field heights.

Further measurements of magnetic fields along solar jets may shed light on this suggestion.

**Data availability**

The data underlying this article will be shared on reasonable request to the corresponding author.